# Quantum Optical Induced-Coherence Tomography by a Hybrid Interferometer

Eun Mi Kim[1,2], Sun Kyung Lee[1], Sang Min Lee[1], Myeong Soo Kang[2], and Hee Su Park[1*]

[1]*Korea Research Institute of Stand and Science (KRISS), Daejeon 34113, South Korea*
[2]*Department of Physics, Korea Advanced Institute of Science and Technology (KAIST), Daejeon 34141, South Korea*
[*]*hspark@kriss.re.kr*

**Abstract**

Quantum interferometry based on induced-coherence phenomena has demonstrated the possibility of undetected-photon measurements. Perturbation in the optical path of probe photons can be detected by interference signals generated by quantum mechanically correlated twin photons propagating through a different path, possibly at a different wavelength. To the best of our knowledge, this work demonstrates for the first time a hybrid-type induced-coherence interferometer that incorporates a Mach-Zehnder-type interferometer for visible photons and a Michelson-type interferometer for infrared photons, based on double-pass pumped spontaneous parametric down-conversion. This configuration enables infrared optical measurements via the detection of near-visible photons and provides methods for characterizing the quality of measurements by identifying photon pairs of different origins. The results verify that the induced-coherence interference visibility is approximately the same as the heralding efficiencies between twin photons along the relevant spatial modes. Applications to both time-domain and frequency-domain quantum-optical induced-coherence tomography for three-dimensional test structures are demonstrated. The results prove the feasibility of practical undetected-photon sensing and imaging techniques based on the presented structure.

## Introduction

Photonic quantum sensing and imaging technologies based on entanglement offer unique possibilities [1-5]. Examples include spatial super-resolution beating the classical limit [3] and photo-detection sensitivity below the standard quantum limit [4]. Moreover, a set of schemes including this work, classified as ghost imaging in a broad sense [2], can separate the probe light in contact with samples and the detection light carrying the measurement information to a detector. In cases where the mere existence of detection photons can infer the correlation between probe-detection photon pairs, the detection of probe photons can even be neglected [5]. Applying this technique to non-degenerate entangled pairs can overcome the optical measurement limitation caused by the intrinsic quantum efficiency of detector materials in specific wavelength bands because the detection wavelength of light can be shifted to fit the available high-quality detectors.

These undetected-photon sensing and imaging techniques rely on the quantum correlation between photon pairs generated by a coherent combination of multiple photon-pair sources. Their archetypal implementation includes probe (*idler*) photons lying on the overlapped path of two sources where the photons can be generated only by the constructive interference of the two generation possibilities. Their twin detection (*signal*) photons are projected onto a coherent superposition state of distinct signal modes from the two sources [6,7]. Therefore, changes in the probe photon path by a target are measurable with interferometry for the signal photons, and the probe photons need not be measured eventually because their existence is guaranteed through energy conservation. This induced-coherence phenomenon, observed since 1991 [8,9], has led to the demonstration of optical measurement techniques with ever-increasing complexity, such as optical on-off imaging [5], spectroscopy [10,11], optical coherence tomography (OCT) [12,13], microscopy [14,15], Fourier-transform infrared spectroscopy [16,17] and holography [18].

A major advantage of applying the induced-coherence technique to OCT is that the non-destructive testing of structures requiring low-energy photons is facilitated by the detection of higher-energy photons that minimize noise. Infrared (IR) optical depth profiling based on time-domain (TD) OCT [12] and wavelength-domain OCT [13, 19] has been demonstrated based on classical types of measurements in the visible wavelength range through induced coherence. These previous quantum optical induced-coherence tomography (QICT) experiments used Michelson-type interferometers, where all the pump, signal, and idler paths passed twice through a single spontaneous parametric down-conversion (SPDC) crystal by the respective back-reflecting mirrors. In contrast, our work constructs a hybrid-type interferometer in which a pump laser and a probe photon propagate through similar Michelson interferometers, while a detection photon path comprises a Mach-Zehnder-type interferometer. We show that this structure, explained theoretically in a review article [20] and realized for the first time to our knowledge, allows us to characterize the heralding efficiencies between the twin photons and hence correctly anticipate the interference visibility. We experimentally verify that the interference visibility, which is critical in maximizing the signal-to-noise ratio (SNR) of QICT, is approximately the average of the heralding efficiencies from signal to idler photons.

The interferometer setup is applied to frequency-domain (FD) QICT in the IR wavelength range of approximately 1.55 μm, through spectral measurement of visible photons near wavelength of 810 nm. The depth profiles of reference samples made of sapphire, silicon, and glass plates are reconstructed by Fourier transform of the spectral interference fringes with all the optical parts fixed. The test patterns of a metal-coated glass plate under a silicon plate cover that generally blocks the detection wavelength (~ 810 nm) are imaged by relative transverse scanning of the probe and sample.

## Principle

We analyze induced-coherence interferometry using a simplified schematic model shown in figure 1. Two SPDC crystals ($DC_1$ and $DC_2$) produce a major part of the signal and idler photons along the spatial modes denoted by $s_n$ and $i_n$ ($n$ = 1,2). Modes $t_n$'s and $j_n$'s are loss channels containing the photons that are not coupled to the major modes. Although such unpaired or stray photons generally scatter into multiple modes, regarding them as single modes is sufficient for the current analysis. The generated photon pairs are initially in the following state:

$$|\psi\rangle = \left[ C_1(p_1\hat{s}_1\hat{i}_1 + q_1\hat{s}_1\hat{j}_1 + r_1\hat{t}_1\hat{i}_1) + C_2(e^{i\phi}p_2\hat{s}_2\hat{i}_2 + q_2\hat{s}_2\hat{j}_2 + r_2\hat{t}_2\hat{i}_2) \right]|0\rangle, \quad (1)$$

where $C_{1,2}$, $p_{1,2}$, $q_{1,2}$, and $r_{1,2}$ are complex constants, and $\hat{x}_{1,2}$ is the creation operator for mode $x_{1,2}$: $\hat{x}_{1,2} \equiv \hat{a}^\dagger_{x_{1,2}}$ ($x = s, i, t, j$). $\phi$ is the phase added by the phase shifter to the $s_2$ path in figure 1. The phase accumulated by the idler path between the two crystals can be considered as an offset of $\phi$ without loss of generality. The heralding efficiency $\mu_{s \to i,n}$ ($\mu_{i \to s,n}$) that a signal (idler) photon heralds an idler (signal) photon in the major modes by $DC_n$ ($n$ = 1,2) is expressed as:

$$\mu_{s \to i,n} = \frac{|p_n|^2}{|p_n|^2 + |q_n|^2}, \quad \mu_{i \to s,n} = \frac{|p_n|^2}{|p_n|^2 + |r_n|^2}. \quad (2)$$

Two lossless beam splitters ($BS_2$ and $BS_3$) in the middle of the interferometer arms have variable transmittances $\eta_s$ and $\eta_i$. These BSs are sets of neutral-density filters used in the experiments to verify the validity of the calculations below. Although equation (1) assumes a single spectral/temporal mode for each of the pump, signal, and idler photons, the general conclusions are still valid for photon pairs in frequency-entangled states from cw-laser-pumped SPDC, as shown from early experiments [8]. This is because the interfering signal paths cannot be distinguished by the spectral/temporal measurement of the idler photons when the idler photon paths are overlapped. Unlike other quantum interferometers using multiple photon-pair sources, photon-wise indistinguishability or purity based on a factorable joint spectrum is not required, and moreover, a high level of frequency entanglement is rather desirable for practical sensing applications.

The final state after propagation through all the components is collected as:

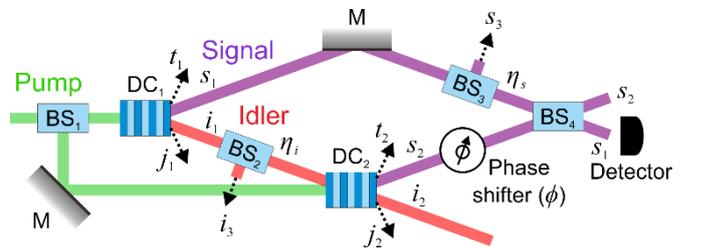

**Figure 1**. Model of the induced-coherence interferometer. Major correlated modes participating in induced-coherence interference are $s_n$ (signal) and $i_n$ (idler), and modes $t_n$ and $j_n$ are occupied by unpaired photons ($n$ = 1,2). DC: down-conversion crystal, BS: beam splitter (1-$\eta_s$, 1-$\eta_i$: reflectance corresponding to added attenuation in experiments), M: mirror.

$$|\psi\rangle = \begin{bmatrix} C_1 \begin{bmatrix} p_1\{\sqrt{\eta_s}(\hat{s}_1+\hat{s}_2)/\sqrt{2}+\sqrt{1-\eta_s}\hat{s}_3\}(\sqrt{\eta_i}\hat{i}_1+\sqrt{1-\eta_i}\hat{i}_3) \\ +q_1\{\sqrt{\eta_s}(\hat{s}_1+\hat{s}_2)/\sqrt{2}+\sqrt{1-\eta_s}\hat{s}_3\}\hat{j}_1 \\ +r_1\hat{t}_1(\sqrt{\eta_i}\hat{i}_1+\sqrt{1-\eta_i}\hat{i}_3) \end{bmatrix} \\ +C_2\begin{bmatrix} p_2 e^{i\phi}(\hat{s}_1-\hat{s}_2)\hat{i}_2/\sqrt{2}+q_2 e^{i\phi}(\hat{s}_1-\hat{s}_2)\hat{j}_2/\sqrt{2} \\ +r_2\hat{t}_2\hat{i}_2 \end{bmatrix} \end{bmatrix}|0\rangle. \quad (3)$$

Tracing out the idler states from the above state after equating the $i_1$ and $i_2$ modes yields the observed signal-photon state. Therefore, the single count rate at the output detector is proportional to:

$$R_s = {}_{s_1}\langle 1|Tr_i\{|\psi\rangle\langle\psi|\}|1\rangle_{s_1} \qquad (4)$$
$$= \frac{1}{2}\left(|C_1p_1|^2\eta_s+|C_2p_2|^2+|C_1q_1|^2\eta_s+|C_2q_2|^2\right)$$
$$+|C_1C_2p_1p_2|\sqrt{\eta_s}\sqrt{\eta_i}\sin(\phi+\phi_0)$$

where $\phi_0$ is a constant phase offset. The visibility $\gamma$ of the final induced-coherence interference fringes is

$$\gamma = \frac{2|C_1C_2p_1p_2|\sqrt{\eta_s}\sqrt{\eta_i}}{(|p_1|^2+|q_1|^2)|C_1|^2\eta_s+(|p_2|^2+|q_2|^2)|C_2|^2}. \quad (5)$$

When the two SPDC processes are symmetric ($|C_1|=|C_2|$, $|p_1|=|p_2|$, $|q_1|=|q_2|$, $|r_1|=|r_2|$), the visibility formula simplifies to $\gamma = \mu_{s\to i}\cdot 2\sqrt{\eta_s}\sqrt{\eta_i}/(1+\eta_s)$. This result signifies the major implication of our analysis that visibility is the same as signal-to-idler heralding efficiency.

For QICT, samples with multiple reflection surfaces are placed in the idler path between the two SPDC crystals shown in figure 1. Each photon has a finite coherence time or bandwidth, and interference appears only when the relevant group delays of signal and idler photons are matched at the detectors. When the pump laser is monochromatic, this group delay matching condition states that the propagation time of the $i_1$-mode photon from DC$_1$ to DC$_2$ ($\tau_0$) must be the same as the propagation time difference between the $s_1$-mode photon from DC$_1$ to BS$_4$ ($\tau_1$) and the $s_2$-mode photon from DC$_2$ to BS$_4$ ($\tau_2$) in figure 1, that is, $\tau_0 - (\tau_1 - \tau_2) = 0$ [9]. When the idler photon path through the sample is fixed and the signal path length is varied, the interference fringes in equation (4) appear only within a delay smaller than the coherence length of the signal photons. Resolving the depth of the layers by scanning the variable delay embodies TD QICT.

The reflection from a sample surface can be modeled as placing $r_1 e^{i\omega_i(2d)/c}$ at the place of $\sqrt{\eta_i}$ in equation (3), where $r_1$ is the amplitude reflection coefficient, $\omega_i$ is the angular frequency of idler photons, and $d$ is the optical thickness (geometrical thicknesses multiplied by refractive indices for relevant sections) with respect to a reference plane. With fixing the lengths of the interferometer arms, interference fringes according to the wavelength ($= 2\pi c/\omega_i$) are the measurement results of the FD QICT. Retaining all the contributions of $s_1$, $s_2$, and $i_1$ modes to the phase of the interference and using the relation that the sum of signal frequency and idler frequency is constant, the magnitude of interference fringe can be expressed as:

$$\gamma\sin(\omega_s(\tau_1-\tau_2)+\omega_i\tau_0+\phi_0) = \gamma\sin(\omega_s(\tau_1-\tau_2-\tau_0)+\phi_0')$$
$$\approx \gamma\sin\left(-(2\pi\Delta(2n_{gi}d_i)/\lambda_{s0}^2)\Delta\lambda_s+\phi_0'\right), \quad (6)$$

where $\lambda_{s0}$ and $\Delta\lambda_s$ are the center wavelength and the relative wavelength of signal photons, respectively, and $\phi_0'$ is a phase constant. $n_{gi}$ and $d_i$ are the group index and the thickness of the sample, respectively. $\Delta(2n_{gi}d_i)$ is the relative optical path length through the idler path, calculated as twice the thickness multiplied by the group index summed over all the sections, and becomes zero at the perfect time delay matching condition ($\tau_0 - \tau_1 + \tau_2 = 0$). When the sample has multiple reflection surfaces, the spectrum of signal photons contains multiple frequency components given by equation (6), and the Fourier transform of the spectrum reconstructs the position and magnitude of the reflection peaks.

## Experimental setup

The experimental configuration used in our work is shown in figure 2. A periodically poled lithium niobate (PPLN) crystal (length 5 mm) is pumped by a continuous-wave laser (wavelength 532 nm) for non-degenerate type-0 SPDC. Photon pairs with signal and idler wavelengths at ~810 nm and ~1550 nm, respectively, are generated along both the forward and backward directions by a pump beam ($1/e^2$-diameter 300 μm) that is focused by a lens (L1, focal length 1000 mm) and refocused by a concave mirror (focal length 200 mm). The forward-propagating signal photons pass through a variable delay line and are superposed with the backward-propagating signal photons at a non-polarizing beam splitter (NPBS), completing a Mach-Zehnder-type interferometer. The forward-propagating idler photons are back-reflected at the sample surface and co-propagate with the backward-propagating idler photons to form a Michelson-type interferometer.

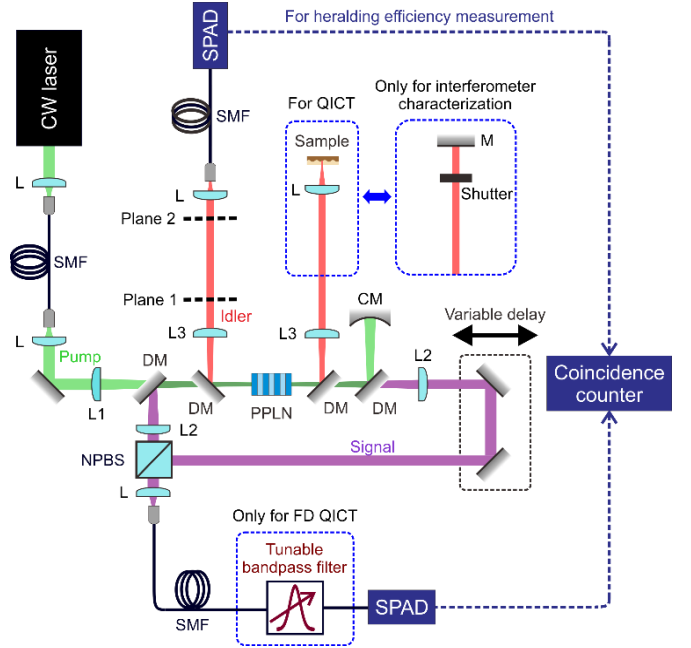

**Figure 2.** Schematic of the experimental setup. A combination of a flat mirror and a shutter is used only for characterization of the interferometer. SMF: single-mode fiber, DM: dichroic mirror, PPLN: periodically poled lithium niobate crystal, CM: concave mirror, L: lens, NPBS: non-polarizing beam splitter, SPAD: single-photon avalanche diode, QICT: quantum-optical induced-coherence tomography, FD: frequency-domain, CW: continuous wave.

The signal photons exiting the NPBS are finally coupled to a single-mode fiber (SMF; Thorlabs 780HP) and detected by a silicon single-photon avalanche diode (SPAD). Although the measurement of idler photons is unnecessary for induced-coherence interferometry, we add a detector for the idler photons to characterize the setup. The spatial intensity distributions of the idler photons heralded by the signal photons are first measured on the transverse planes (planes 1 and 2 in figure 2). An SMF (Corning SMF-28) that best fits the measured profiles is inserted together with a relevant coupling aspheric lens (focal length 11 mm) and connected to an InGaAs SPAD. This geometry enables us to measure all the heralding efficiencies between the relevant signal and idler photons by comparing the coincidence counts and the single counts of the two SPADs by blocking the relevant signal, idler, and pump paths by optical shutters. This feature differentiates this study from the previous studies. Final measurements, including QICT experiments, record only single counts using the silicon SPAD, selectively incorporating bandpass filtering when necessary.

The collimation lenses (L2 and L3 in figure 2) for signal and idler beams have focal lengths of 300 mm and 250 mm, respectively. The aspheric lens collecting the signal photons into the SMF has a focal length of 6.2 mm, therefore the detected signal photons nominally have a $1/e^2$ mode field diameter of 240 μm at the SPDC crystal. To verify the spectral and spatial overlaps of the two photon-pair generation possibilities, we measure the spectrum of signal photons and the spatial profiles of idler photons. Overlapping the spatial modes of light along a shared path inside the SPDC crystal achieves spectral identity determined by the phase-matching condition. Figure 3(a) shows the measured spectra of the forward- and backward-generated signal photons, which are indistinguishable within the experimental uncertainty.

At planes 1 and 2 in figure 2, which were 600 mm apart from each other, the spatial distribution was measured by transverse scanning by a step-index multimode fiber tip (Thorlabs M42L02, core diameter 50 μm) to collect the idler photons, as shown in figure 3(b). The unit step for the scanning was 60 μm along both the x- and y-directions. The measured $1/e^2$ beam diameters were 2.2 mm and 2.9 mm for forward- and backward-pumped idler photons, respectively, and the beam divergences were < 0.1 mrad. The SMF and lens collecting the idler photons nominally match with an incoming beam diameter of 2.5 mm and project a 210-μm diameter spot on the SPDC crystal. Both the forward- and backward-pumped idler photons are aligned individually to maximize the coupling efficiencies to the SMF. Note again that this SMF coupling is independent of the QICT experiments and is used only to verify the relations regarding interference visibility.

The heralding efficiencies among photon pairs are measured with signal and idler photons coupled to the respective SMF cables. After compensating for the detection efficiencies of the SPADs and the back-reflection losses at the fiber end faces, the signal-to-idler (idler-to-signal) heralding efficiency $\eta_{s\to i}$ ($\eta_{i\to s}$) was 63% (43%) for the forward-pumped SPDC. The backward-pumped SPDC showed $\eta_{s\to i}$ ($\eta_{i\to s}$) of 60% (49%). Departure from the ideal 100% efficiency can be attributed to imperfect spatial mode matching, possibly by phase nonuniformity that is not resolvable in the intensity profiles in figure 3(b) and transmission losses of other optical components. Although there is no evidence that the SMF collects the main common idler mode ($i_1$ and $i_2$ in figure 1), we expect that the spatial mode defined by the SMF is a good approximation because the mode field diameter (2.1 mm) is close to that of heralded idler photons (2.2 mm and 2.9 mm), and the heralding efficiencies for forward and backward pumped photon pairs are close to each other.

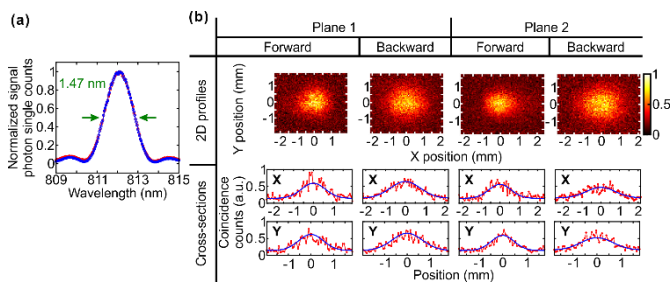

**Figure 3.** Spectral and spatial characteristics of photons. (a) Spectra of forward (red) and backward (blue) pumped signal photons. (b) Spatial intensity distributions of heralded idler photons measured at planes 1 and 2 in figure 2. The blue lines in the cross-sections are Gaussian fit (with constant background) to the measured data (red dots).

### Induced-coherence interference visibility

To investigate the interference visibility, we first place a mirror at the sample position, as shown in figure 2. An interference fringe burst appears according to the position of the signal delay line, as shown in figure 4(a). Figure 4(b) shows the magnified fringes observed by fine-tuning the idler reference mirror position with a piezoelectric actuator with a signal delay fixed at the center of the interference fringe profile. The measured interference visibility was 64%.

The spectral-domain interference fringes in equation (6), with an offset delay of 1 mm in the signal path, are shown in figure 4(c). Here, the average counts in the absence of interference were measured blocking one idler path and subtracted from the measurement data. This subtraction removes an unwanted zero-delay (dc-component) artifact peak in the following results. Figure 4(d) is the Fourier transform of the spectrum in figure 4(c), where the optical path length $\Delta d_s$ is calculated by equating the phase factor in equation (6) to $2\pi\Delta d_s/(\lambda_{s0}\cdot\Delta\lambda_s)$. The calculated value clearly follows the actual position of the signal delay, as shown in figure 4(e).

We note that the measured visibility (64%) is close to the heralding efficiencies $\eta_{s\rightarrow i}$'s (63% and 60%) and verifies the relation in equation (5). We further confirm the relation experimentally by placing neutral-density filters in one of the signal or idler interferometer arms (adjacent to the signal delay line or the sample). Figures 5(a) and 5(b) show the visibility according to the transmission, and confirm the theoretical calculation in equation (5). The linear dependence of the visibility in figure 5(a) can be interpreted as follows: among the detected signal photons, the unpaired signal photons, not heralding idler photons into the major spatial mode, constitute a constant background noise. The ratio of the paired signal photons to the total signal photons is approximately the same as the visibility. Figure 5(b) shows the typical behavior of conventional two-beam interferometers for the variable transmittance of one arm.

## Applications to tomography and imaging

We describe the QICT and imaging configuration. The basic characteristics of the current interferometer setup for optical depth profiling are shown in figure 6. The axial resolution in figure 6(a) is the full width at half-maximum of the peaks in figure 4(d) and inversely proportional to the spectral width of photons. For FD QICT, peaks in the positive and negative optical delays are overlapped and not distinguished. This aliasing effect narrows the peaks with near-zero optical delay, as shown in figure 6(a). During measurements of real samples, we avoid the aliasing ambiguity by locating a reference position such that all the reflection surfaces lie at positive or negative delay positions. The average half-maximum resolution was 0.194 mm, which agrees with the theoretical estimation (0.198 mm) from the spectrum in figure 3(a) [21]. The range of measurable optical depth was determined by the resolution (0.07 nm) of spectral measurements. The sensitivity measured by the amplitude of the spectral peak in figure 4(d) decreases according to the optical delay, as shown in figure 6(b) [22]. The SNR is proportional to the integration time for unit measurement, as shown in figure 6(c), which indicates that the noise is mainly governed by the Poissonian photon number fluctuation (shot noise) of photon counting.

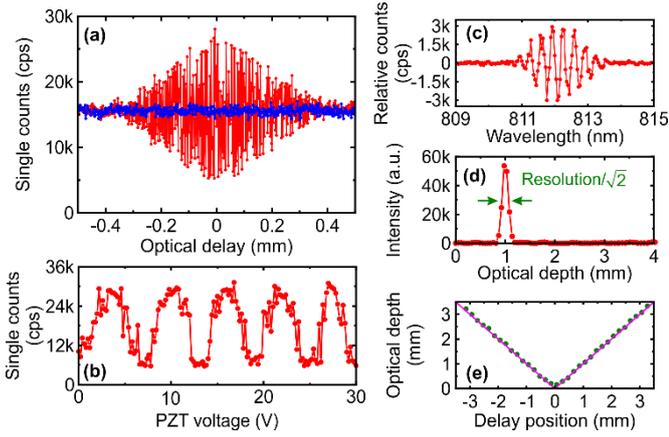

**Figure 4.** Quantum optical induced-coherence interference fringes in the time and frequency domains. (a) Interference fringes according to the optical delay for signal photons without (red) and with (blue) blocking the idler path between the two crystals. (b) Fine-scanning results using piezoelectric control of the mirror position at the idler path. (c) Interference fringe in the spectral domain with the optical delay for signal photons fixed at 1 mm that is greater than the coherence length. (d) Optical depth profile derived from the Fourier-transform of (c). (The depth resolution defined as the overall profile width in (a) corresponds to the width of amplitude profile, hence an additional factor of $1/\sqrt{2}$.) (e) Measured depth (green dots) of the variable delay with respect to the zero-delay position determined in (d). The lines plot the absolute values.

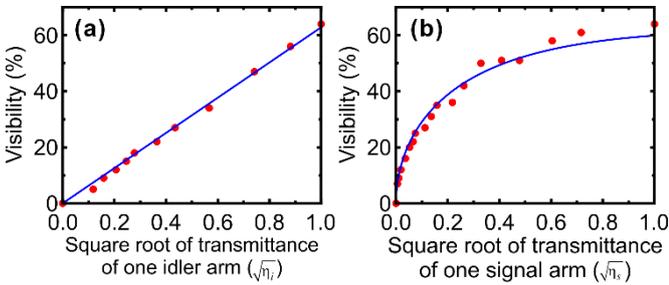

**Figure 5.** Interference visibility according to the transmission loss of (a) one idler arm and (b) one signal arm shown in figure 1. The red dots are experimental data, and the blue lines are theoretical predictions.

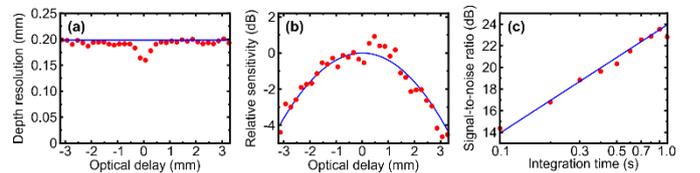

**Figure 6.** Tomography characteristics of the induced-coherence interferometer. (a) Depth resolution according to the optical delay. Small reduction at the zero delay is caused by the aliasing effect of Fourier transform. (b) Relative sensitivity according to the optical delay. (c) Signal-to-noise ratio with the integration time. Red dots: experiments, blue lines: theory.

Two types of reference samples were applied for depth profiling. Sample 1 has three reflection planes from sapphire-air-silicon layers within the detection range, as shown in figure 7(a). The TD and FD QICT results are shown in figures 7(b) and 7(c), respectively. Considering the group index of sapphire (1.77) at the idler wavelength (1550 nm), the thicknesses of the air and sapphire layers corresponding to the results in figure 7(c) are 0.431 mm and 0.442 mm, respectively, and are within 1% of the values measured from the microscope image in figure 7(a). Sample 2 is a combination of a double-side-polished silicon plate and a sapphire plate and has four major reflection components including boundaries with air and double reflection within the highly reflective silicon, as shown in figure 7(d). The TD and FD QICT results in figures 7(e) and 7(f) clearly resolve these four peaks. The group index of silicon is 3.61, and the thicknesses of silicon and sapphire layers estimated in figure 7(f) were 0.251 mm and 0.489 mm, respectively, which agreed within 8% with the microscope image. The additional peaks appearing in the FD QICT can be attributed to unidentified multiple reflections or interference among them. For example, the peak at near-zero depth corresponds to a path composed of double roundtrips in both the silicon and sapphire plates. The peak between cases c and case d (near an optical depth of 1 mm) can be attributed to double roundtrips inside the silicon plate plus triple roundtrips inside the sapphire plate. Other peaks can also appear owing to interference between different reflection components within the sample path when the sample surfaces are highly reflective. Such peaks can be removed from the detection range by placing the reference plane on the other side of the sample. (The current reference position was chosen to better resolve the weakly reflecting cases.)

To demonstrate the imaging capability for targets that are normally invisible at visible wavelengths, we prepared a resolution test target (1951 USAF, Thorlabs R3L3S1N) patterned on a chrome-coated soda-lime glass substrate covered with a polished silicon wafer, as shown in figure 8(a). A lens (focal length of 35 mm) focuses the idler photons onto the sample and re-collimates the reflected photons. The half-maximum spot diameter was estimated to be 17 μm (20 μm) for the x (y) direction considering the measured idler mode size (figure 3(b)). Measuring the edge response at the target yields the line spread function, from which the transverse resolution was determined as 11.2 μm (13.2 μm) along the x (y) direction, as shown in figure 8(b). The resolution is smaller than the spot size by approximately $\sqrt{2}$ because when a sample reflects a fraction $f$ of the incident transverse beam profile, only $f^2$ of the beam returns to the collecting mode because of reduced mode overlap. The slight asymmetry can be attributed to the residual asymmetry of our setup along the horizontal and vertical directions of the SPDC configuration.

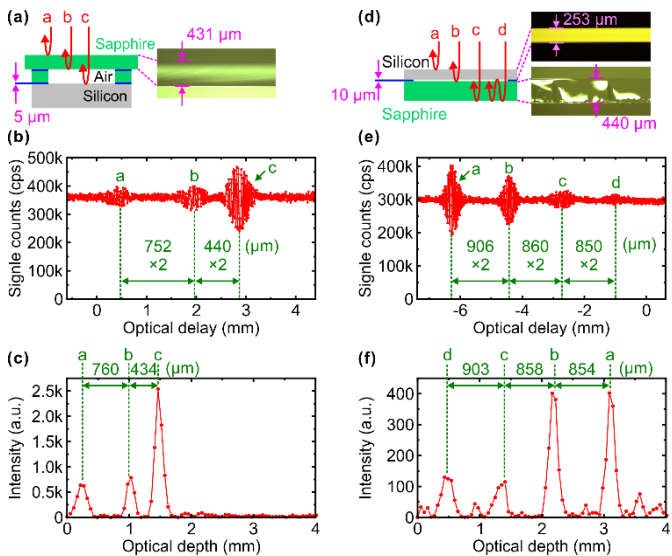

**Figure 7.** Depth profiling of test samples. (a) Structure of sample 1. (b) TD QICT and (c) FD QICT measurements for sample 1. (d) Structure of sample 2. (e) TD QICT and (f) FD QICT measurements for sample 2.

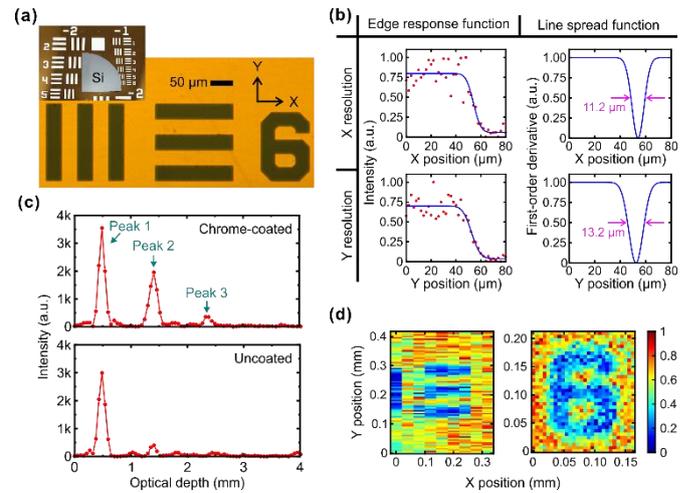

**Figure 8.** Transverse imaging with FD QICT. (a) Optical micrograph of a 1951 USAF resolution target with negative patterns on a chrome-coated glass. The target is covered with a silicon plate that is normally opaque for visible light, as shown in the inset photograph. (b) Transverse resolution for FD QICT. Edge response functions are measured (red dots) and fitted to error functions (blue lines). The line spread functions as their derivatives lead to transverse resolution along the x- (y-) direction of 11.2 (13.2) μm. (c) Optical depth profile of the reflected idler photons. Peaks 1 and 2 show reflections at the outer surface of the silicon plate and at the chrome coating, respectively. Peak 3 arises from multiple reflections between the silicon plate and the chrome coating. (d) Transverse image measured through the silicon cover plate.

FD QICT obtains images with relative transverse scanning of the lens and the sample. Figure 8(c) compares the axial optical-depth profiles for the cases where the focused idler photons impinge on either the uncoated patterns or the coated background. Peaks 1 and 2 correspond to the major reflections at the silicon plate and the chrome coating, respectively, and peak 3 is caused by multiple reflections at the two surfaces. The uncoated area shows a smaller reflection at the silicon-glass interface at the peak 2 position. Extracting the signal intensity distribution at the peak 2 position generates the images shown in figure 8(d).

## Conclusion and outlook

In summary, this work has verified the feasibility of a hybrid-type induced-coherence interferometer for applications to IR optical tomography based on the detection of near-visible photons. The demonstrated configuration provides versatile independent control over the signal and idler paths, thereby matching any idler-path geometry designed for samples by adjusting the signal delay length. We believe that this feature opens up possibilities for applications to other sensing technologies, such as spectroscopic gas sensing, retaining the advantage of undetected-photon measurements. The theoretical analysis and the experimental verification have emphasized the importance of heralding efficiencies between photon pairs in terms of the induced-coherence interference visibility. Both the TD and FD QICT experiments have successfully reconstructed the three-dimensional structures of test samples.

Regarding further improvement of the performances of optical tomography, first, the depth-profiling resolution can be sharpened by increasing the wavelength bandwidth satisfying the phase matching condition. A shorter-length SPDC crystal and/or tighter focusing of a pump beam are straightforward directions because they reduce the magnitude of phase mismatch for a given wavelength and increase the chance of phase matching by widening the range of longitudinal wavenumber. Photon-pair generation efficiencies and heralding efficiencies must be considered for trade-offs. An aperiodically poled crystal has also been used to realize ultra-broadband SPDC in a highly non-degenerate configuration [23]. Secondly, the data acquisition time and the overall range of depth profiling depend on the measurement capabilities. Single-photon-level spectrometers based on an SPAD array [24] instead of a single SPAD with a scanning bandpass filter can reduce the measurement time. Narrowing the bandwidth of unit spectral measurement increases the depth range of tomography. As the induced-coherence technique is generally compatible with various existing optical sensing technologies, incorporating advanced OCT techniques for wavelength-swept light sources and signal processing routines is worth considering to further enhance the practical advantage of QICT.

## Acknoledgments

This research was supported by the National Research Council of Science & Technology(NST) grant (CAP22051-000), the KRISS project (GP2023-0013-07), and the education and training program of the Quantum Information Research Support Center funded through the NRF, all by the Korean government (MSIT). We thank Dongha Kim of KAIST for the help in preparing the measurement samples.